\documentclass[12pt]{JHEP3}

\usepackage{epsfig,multicol} 

\makeatletter
\def\eqnarray{%
 \stepcounter{equation}%
 \let\@currentlabel=\theequation
 \global\@eqnswtrue
 \global\@eqcnt\z@
 \tabskip\@centering
 \let\\=\@eqncr
 $$\halign to \displaywidth\bgroup\@eqnsel\hskip\@centering
 $\displaystyle\tabskip\z@{##}$&\global\@eqcnt\@ne
 \hfil$\displaystyle{{}##{}}$\hfil
 &\global\@eqcnt\tw@$\displaystyle\tabskip\z@{##}$\hfil
 \tabskip\@centering&\llap{##}\tabskip\z@\cr}
\makeatother

\title{ The Disc Amplitude  of 
    the Dijkgraaf-Vafa Theory:1/N Expansion vs Complex Curve Analysis }

\author{Shogo Aoyama\\
        Department of Physics, Shizuoka University, 
          Ohya 836, Shizuoka, Japan  \\
        E-mail: \email{spsaoya@ipc.shizuoka.ac.jp}}

\abstract{According to Dijkgraaf and Vafa the effective glueball superpotential of the ${\cal N}=1$ supersymmetric QCD coupled with an adjoint chiral multiplet  is given  by  the planar amplitude in the ${1\over N}$ expansion of a matrix model.  It was shown  that, when the ${\cal N}=1$ supersymmetric QCD is coupled with fundamental chiral multiplets as well, the effective glueball superpotential is modified by the disc amplitude of the  generalized matrix model. The diagramatic computation of this disc amplitude is fairly involved for the multi-cut solution. Instead we compute it with recourse to the complex analysis of the hyperelliptic curve. The result is given  in series of the gluino condensation $S_i $. 
The explicit computation for the generic multi-cut solution is done up to order $S^3$. It is systematic so that  it can be  extended to higher orders. }

\keywords{Matrix Models, Supersymmetric Effective Theories, Nonperturbative Effects, {1/N} Expansion}

\preprint{}

\begin{document}

\section{Introduction}

Dijkgraaf and Vafa discovered a close relation between the ${\cal N}=1$ supersymmetric QCD and a matrix model. They claimed that the 
glueball superpotential of the ${\cal N}=1$ supersymmetric QCD coupled with an adjoint chiral multiplet can be computed by the ${1\over N}$ expansion of an associated matrix model\cite{DV}.  It was confirmed by subsequent works by many people. 
The effective theory of the ${\cal N}=1$ supersymmetric QCD  given by such a glueball superptential  is called the Dijkgraaf-Vafa (DV) theory. 
In \cite{DV} they developed a remarkable technique to compute the planar diagram amplitude of the matrix model. Namely the computation, which was normally done  by the perturabtive sum of the planar  diagrams, was equivalently replaced by the analysis of the hyperelliptic curve on the Riemann surface. This revealed important aspects underlying in the ${\cal N}=1$ supersymmetric QCD like the duality\cite{Cacha}, the integrable hierarchy, {\it etc}, and attracted a renewed  attention\cite{renew,Morosov} to the Seiberg-Witten theory for the $N=2$ supersymmetric QCD\cite{SW}.

In \cite{Ar, HS, Sch, Wit} the relation between the  the ${\cal N}=1$ supersymmetric QCD and a matrix model was generalized for the case in which  the ${\cal N}=1$ supersymmetric QCD is coupled with fundamental chiral multiplets as well. The presence of  fundamental chiral multiplets modifies the glueball superpotential of the ${\cal N}=1$ supersymmetric QCD.  
It was shown  that the modification can be captured by computing the planar diagrams with boundary (discs)  of  the matrix model which is generalized accordingly. The disc amplitude is the perturbative sum of such disc diagrams. It was computed  to all orders of the gluino condensation $S$ for the $1$-cut solution\cite{Ar}, but merely to order $S^2$ for the $n$-cut solution\cite{Sch}. The computation to higher orders  for the $n$-cut solution is extremely involved. In this paper we will discuss that the disc amplitude  for the $n$-cut solution may be computed to higher orders in a rather systematic way,  if one has  
 recourse to the complex anlysis of the hyperelliptic curve. Indeed the planar amplitude for the $2$-cut solution was calculated along this line to  order $S^3$  in \cite{DV2}. This calculation to  order $S^3$ was generalized for the $n$-cut solution  in \cite{Ito} later on. Our aim of this paper is to generalize  the calculation of the planar amplitude  in \cite{Ito} to that of  the disc amplitude.

The paper is organized as follows. We give a short review on the DV theory coupled with fundamental chiral multiplets in section 2. The disc amplitude of the associated matrix model is given in terms of the (hyper)elliptic function. 
As a warming-up it is  evaluated for the $1$-cut solution in section 3. The evaluation is exact for this case. For the multi-cut solution it 
  can be done   only in  series of  the gluino condensation $S$. 
In section 4   the disc amplitude for the $2$-cut solution  is evaluated  to order $S^3$.  In section 5 this evaluation is generalized for the $n$-cut solution.

\section{A Short Review on the DV theory}

We consider the ${\cal N}=1$ supersymmetric $U(N)$ gauge theory with the superpotential\begin{eqnarray}
W(\Phi,Q^I,\tilde Q_I) = tr V(\Phi) + \sum^{N_f}_{I=1} 
(m_I\tilde Q_I Q^I + Q_I \Phi Q^I),    \label{superpot}
\end{eqnarray}
Here $\Phi$ is a adoint chiral multiplet and $Q^I(\tilde Q_I)$ are $N_f$ chiral multiplets in the fundamental representation. $V(\Phi)$ is a polynomial scalar potential of  degree $M_0$ given by 
\begin{eqnarray}
V'(x) = (x-\alpha_1)(x-\alpha_2)\cdots(x-\alpha_{M_0}).   \label{genepot}
\end{eqnarray}
When the $U(N)$ gauge group symmetry is broken to $\Pi_i U(N_i)$ with $\sum^n_{i=1} =N$, this gauge theory is described by the effective action 
\begin{eqnarray}
W_{eff}= \sum_{i=1}^n N_i{\partial F_2(S)\over \partial S_i} + F_1(S)
 + 2\pi i \tau_0\sum_{i=1}^n S_i, \label{effective}
\end{eqnarray}
for the gluino condensation $S_i$. It was claimed that $F_2(S)$ and $F_1(S)$ are the planar and disc amplitudes in the ${1\over M}$ expansion of the partition function of the associated matrix model
\begin{eqnarray}
Z &=& \int d\phi dq^I d\tilde q_I \exp \Bigl[-{1\over g_s}W(\phi,q^I,\tilde q_I)\Bigr]  \nonumber \\
 &=& \exp\Bigl[-\sum_{\chi \le 2} g_s^{-\chi}F_\chi (S) \Bigr]. \nonumber
\end{eqnarray}
Here $\phi$ is a $M\times M$ matrix and $q^I(\tilde q^I)$ are $M$ dimensional vectors. $W$ is the superpotential given by (\ref{superpot}) so that the matrix model is gauge-invariant under $U(M)$. The gluino condensation $S_i$ is identified with the 'tHooft coupling $g_s M_i$, when the gauge symmetry $U(M)$ is broken to $\Pi_i U(M_i)$ with $\sum^n_{i=1} =M$. 
The planar and disc amplitudes, respectively given by $F_2$ and $F_1$,  are calculated by an alternative method. Namely we think of 
the bosonic matrix model
\begin{eqnarray}
Z &=& \int d\phi \exp \Bigl[-{1\over g_s}tr V(\phi)\Bigr]. \nonumber 
\end{eqnarray}
Then we obtain the saddle point equation for the eigenvalues distributed with density $\varrho(\lambda)$
\begin{eqnarray}
y \equiv  V'(x) + 2\int d\lambda {\varrho(\lambda)\over x-\lambda } =0. \label{SPE}
\end{eqnarray}
Here $y$ is the quantum deformed force of the  matrix model. 
For the $n$-cut solution it may be written as a hyperelliptic function 
 such that
\begin{eqnarray}
y = (x-\mu_1)(x-\mu_1)\cdots (x-\mu_{M_0-n}) \sqrt{(x-\lambda_1)(x-\lambda_2)\cdots (x-\lambda_{2n})}.  \label{hyper}
\end{eqnarray}
Comparing the polynomial part in the expansion of (\ref{SPE}) and (\ref{hyper}) at $x=\infty$ gives the zero-points $\alpha_a, a=1,2,\cdots,M_0$ as functions of $\mu$'s and $\lambda$'s. We consider the DV differential $ydx$ defined with such a  hyperelliptic function.  For the $n$-cut solution the gluino condensation is given by 
\begin{eqnarray}
S_i ={1\over 4\pi i} \int_{A_i} ydx.   \label{S}
\end{eqnarray}
The planar amplitude\cite{DV} was given by 
\begin{eqnarray}
{\partial F_2 \over \partial S_i} = \int^\Lambda_{\alpha_i} ydx,
\end{eqnarray}
while 
the disc amplitude\cite{Sch,Wit}  by 
\begin{eqnarray}
F_1 = {1\over 2}\sum_{I=1}^{N_f}\Bigl[ \int_{-m_I}^\Lambda ydx - V(\Lambda) +V(-m_I) \Bigr],
 \label{disc}
\end{eqnarray}
with a cut-off parameter $\Lambda$. 

We want to calculate this disc amplitude for the $n$-cut solution and compare it with the result obtained by the ${1\over M}$ expansion. For simplicity  we shall do it  simply assuming  that $M_0 = n$.

\vspace{0.5cm}

\section{The $1$-cut Solution}

As a warming-up we start with the simplest example, {\it i.e.}, the matrix model with the scalar potential 
$$
V(x) = {1\over 2}\xi x^2,
$$
which is coupled with one chiral matter multiplet. 
We consider the DV differential 
\begin{eqnarray}
ydx = \sqrt{V'(x)^2 - 4\xi S}dx,  \nonumber
\end{eqnarray}
on the Riemann surface with one branch-cut. Note that the branch points are chosen so that (\ref{S}) is satisfied, {\it i.e.}, 
\begin{eqnarray}
S = {1\over 2\pi i} \mathop{\int}^{\sqrt{{4S\over \xi}}}_{\sqrt{-{4S\over \xi}}} ydx.   \nonumber
\end{eqnarray}
The disc amplitude (\ref{disc}) is exactly calculated as
\begin{eqnarray}
F_1(S) = -{S\over 2} - {1\over 4\alpha}(\sqrt{1-4\alpha S}-1) 
+S\log {1 + \sqrt{1-4\alpha S} \over 2} + S\log {m\over \Lambda},  \nonumber
\end{eqnarray}
with $\alpha = {1\over m^2\xi}$. 
This is in complete agreement with the result obtained by the ${1\over N}$ expansion\cite{Ar}. The log-term which is linear in $S$ is a non-perturbative part and  may be 
absorbed by the last term in the effective action (\ref{effective}). A similar
 calculation for the $1$-cut solution was done for a variant of the matrix model  in \cite{Wata}. 

\vspace{0.5cm}

\section{The $2$-cut Solution}

Next we study the matrix model in which the scalar potential is given by 
\begin{eqnarray}
V'(x) = (x-\alpha_1)(x-\alpha_2).  \label{pot}
\end{eqnarray}
In \cite{Sch} they calculated Feynmann diagrams dipicted in Fig. 1 to find the disc amplitude to  order $S^2$:\FIGURE{\epsfig{file=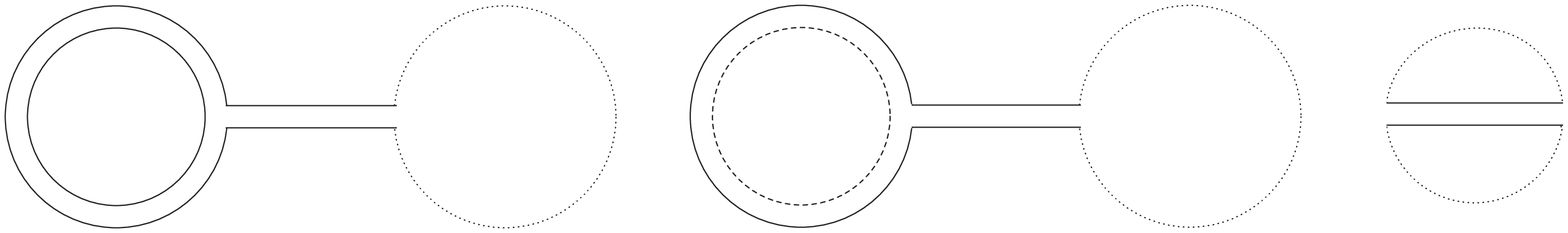, width=13cm}
\caption{The Feynman diagrams to order $S^2$ corresponding to the three terms in (\ref{amp}). The solid double line and the single dashed line  are  propagators for $\phi$ and  $q^I$ respectively. The solid and dashed double line is a propagator for ghosts.}}
\begin{eqnarray}
F_1(S) = -\sum_{I=1}^{N_f} \sum_{i=1}^2 \Bigl[{S_i^2 \over \Delta_i(\alpha) f_{iI}}  {1\over \alpha_{ij}}
-2 \sum_{j\ne i}^2 {S_i S_j \over \Delta_i(\alpha) \alpha_{ij} f_{iI} } +{1\over 2}{S_i^2\over \Delta_i(\alpha) f^2_{iI}}  \Bigr].  \label{amp}
\end{eqnarray}
Here $\Delta_i(\alpha) = {\scriptstyle{\mathop{\prod}}_{j\ne i}^2 }\alpha_{ij}, f_{iI}= m_I + \alpha_i, \alpha_{ij} = \alpha_i - \alpha_j$. In this section we shall evaluate the formula (\ref{disc}) for the $2$-cut solution to order $S^3$ and show that it indeed reproduces the  result (\ref{amp}) to order $S^2$. To this end we consider the DV differential 
\begin{eqnarray}
ydx = \sqrt{\Bigl((x-\gamma_1)^2-\rho_1^2\Bigr)\Bigl((x-\gamma_2)^2-\rho_2^2\Bigr)}dx, \label{diff1}
\end{eqnarray}
on the Riemann surface with two branch-cuts $[\gamma_1-\alpha_1,\gamma_1+\alpha_1]$ and  $[\gamma_2-\alpha_2,\gamma_2+\alpha_2]$. Here $y$ may be written in the form (\ref{SPE}). 
 Hence comparing the polynomial part in  the expansion  of (\ref{SPE}) and (\ref{diff1})   at $x= \infty$ yields the relations $\alpha_i(\rho^2,\gamma)$:
\begin{eqnarray}
\alpha_1 + \alpha_2 &=& \gamma_1 + \gamma_2,  \label{alpha1}  \\
(\alpha_1 + \alpha_2)^2 + 2\alpha_1\alpha_2 &=& \gamma_1^2 +\gamma_2^2 +4\gamma_1\gamma_2 -\rho^2_1-\rho^2_2. \label{alpha2}
\end{eqnarray}
With the DV differential (\ref{diff1}) we will calculate the disc amplitude (\ref{disc}). A suitable technique to this end was developed for the calculation of the planar amplitude in \cite{Ito}. The calculation goes following the steps:

\begin{enumerate}
\item evaluation of the period integrals 
\begin{eqnarray}
S_i = {1\over 2\pi i}\int_{\gamma_i-\rho_i}^{\gamma_i+\rho_i} ydx, 
\label{SS}
\end{eqnarray}
which provides  explicit formulae $S_i(\rho^2,\gamma)$ for gluino condensation.

\item inversion of the formula $S_i(\rho^2,\gamma)$ which provides  $\rho_i^2(S,\gamma) $ in series of $S_i$ with  coefficients made out of $\gamma_i$.

\item inversion of the formulae $\alpha_i(\rho^2,\gamma)$, given by (\ref{alpha1}) and (\ref{alpha2}),  
 which provides  $\gamma_i(\rho^2,\alpha)$ in series of $\rho^2_i$ with coefficients made out of $\alpha_i$. 
\item evaluation of the DV differential (\ref{diff1}) in series of $\rho^2_i$ with coefficients made out of $\gamma_i^2$

\item expansion of $\rho_i^2$ and $\gamma_i$ in the above result in series of $S_i$ with the coefficients made out of $\alpha_i$ by iterating  the inversion formulae obtained at Step 2 and 3. 
\end{enumerate}

\noindent
\underline{Step 1}

After the shift $x'=x-\gamma_1$ we expand the DV differential (\ref{diff1}) in series of $\rho^2_2$ as 
\begin{eqnarray}
ydx = \sum_{m=0}^\infty c_m\rho_2^{2m}(x'+\gamma_{12})^{1-2m}\sqrt{x'^2-\rho^2_1}dx',
\end{eqnarray}
where $\gamma_{12}=\gamma_1-\gamma_2$ and $c_m= {\Gamma(m-{1\over 2})\over m!\Gamma(-{1\over 2})}$. 
Write $(x'+\gamma_{12})^{1-2m}$ for $ m\ge 1$ as $(2m-2)$-ple derivative of $(x'+\gamma_{12})^{-1}$ by $\gamma_{12}$. Then it takes the form 
\begin{eqnarray}
ydx = \Bigl[(x'+\gamma_{12})\sqrt{x'^2-\rho_1^2} + \sum_{m\ge 1}{c_m\over (2m-2)!}\rho_2^{2m}({\partial \over \partial \gamma_{12}})^{2m-2}{\sqrt{x'^2-\rho_1^2}\over x'+\gamma_{12} }\Bigr]dx'.   \label{DV}
\end{eqnarray}
We integrate the DV differential of this form along the $A_1$-cycle surrounding the branch-cuts $[\gamma_1-\rho_1,\gamma_1+\rho_1]$. We find
\begin{eqnarray}
i\int^{\rho_1}_{-\rho_1}dx' (x'+\gamma_{12})\sqrt{\rho_1^2-x'^2}
 = {i\pi\over 2}\gamma_{12} \rho_1^2,
\end{eqnarray}
for the first piece and 
\begin{eqnarray}
i\int^{\rho_1}_{-\rho_1}dx'{\sqrt{\rho_1^2-x'^2}\over x'+\gamma_{12}}
= i\pi(-\sqrt{\gamma_{12}^2 -\rho_1^2} + \gamma_{12}),
\end{eqnarray}
for the second piece. Putting them together in (\ref{SS}) then we get the expression for the gluino condensation $S_1$
\begin{eqnarray}
S_1 = {1\over 4}\gamma_{12}\rho_1^2 -{\rho_1^2\rho_2^2\over 8\gamma_{12}}-
  {\rho_1^2\rho_2^2 (\rho_1^2 + \rho_2^2) \over 32\gamma_{12}^3}+   O(\rho^8).
 \label{S1}
\end{eqnarray}
Similarly we get for the gluino condensation $S_2$
\begin{eqnarray}
S_2 = -{1\over 4}\gamma_{12}\rho_2^2 +{\rho_1^2\rho_2^2\over 8\gamma_{12}}+ 
 {\rho_1^2\rho_2^2 (\rho_1^2 + \rho_2^2) \over 32\gamma_{12}^3}+  O(\rho^8).
 \label{S2}
\end{eqnarray}

\noindent
\underline{Step 2}

Inverting the relations (\ref{S1}) and (\ref{S2}) we obtain
\begin{eqnarray}
\rho_1^2 &=& {4S_1\over \gamma_{12}} - {8S_1S_2\over \gamma_{12}^4} -
  {24S_1S_2(S_1-S_2)\over \gamma_{12}^7} +  O(S^4),
  \label{rho1} \\
\rho_2^2 &=& -{4S_2\over \gamma_{12}} - {8S_1S_2\over \gamma_{12}^4} - 
 {24S_1S_2(S_1-S_2)\over \gamma_{12}^7} +       O(S^4).
 \label{rho}
\end{eqnarray}

\noindent
\underline{Step 3}

Inverting the relations (\ref{alpha1}) and (\ref{alpha2}) we obtain
\begin{eqnarray}
\gamma_1 &=& {1\over 2}(\alpha_1+\alpha_2)+{1\over 2}\sqrt{(\alpha_1-\alpha_2)^2-2(\rho_1^2+\rho_2^2)},  \nonumber \\
 &=& \alpha_1 - {\rho_1^2+\rho_2^2\over 2\alpha_{12}} -{(\rho_1^2+\rho_2^2)^2\over 4\alpha_{12}^3}+ O(\rho^6),  \label{gamma1} \\
\gamma_2 &=& {1\over 2}(\alpha_1+\alpha_2)-{1\over 2}\sqrt{(\alpha_1-\alpha_2)^2-2(\rho_1^2+\rho_2^2)}  \nonumber \\
&=& \alpha_2 + {\rho_1^2+\rho_2^2\over 2\alpha_{12}} +{(\rho_1^2+\rho_2^2)^2\over 4\alpha_{12}^3}+ O(\rho^6). \label{gamma}
\end{eqnarray}

\noindent
\underline{Step 4}

We expand the DV differential (\ref{DV}) in series of $\rho^2_1$ as well. 
\begin{eqnarray}
ydx&=& \Bigl[ V'(x) - {\gamma_{12}\over 2}({\rho_1^2\over x'}-{\rho_2^2\over x'+\gamma_{12}})     
 - {1\over 8}\{{\rho_1^4\over x'^2}+{\rho_2^4\over (x'+\gamma_{12})^2}\}
\nonumber \\
  & -& {\gamma_{12}\over 8}\{{\rho_1^4\over x'^3}-{\rho_2^4\over (x'+\gamma_{12})^3}\}
+{\rho_1^2\rho_2^2 \over 4\gamma_{12}}({1\over x'}-{1\over x'+\gamma_{12}}) 
 \nonumber \\
&-& {1\over 16}\{{\rho_1^6\over x'^4}+{\rho_2^6\over (x'+\gamma_{12})^4}\}
 - {\gamma_{12}\over 16}\{{\rho_1^6\over x'^5}-{\rho_2^6\over (x'+\gamma_{12})^5}\}    \label{DV'} \\
&+& {1\over 16\gamma_{12}}\{{\rho_1^4\rho_2^2\over x'^3}-{\rho_1^2\rho_2^4\over (x'+\gamma_{12})^3}\}  -{1\over 16\gamma_{12}^2}\{{\rho_1^4\rho_2^2\over x'^2}+{\rho_1^2\rho_2^4\over (x'+\gamma_{12})^2}\}  \nonumber \\
&+& {\rho_1^2\rho_2^2 (\rho_1^2 + \rho_2^2)\over 16\gamma_{12}^3}({1 \over x'}  - {1 \over x'+ \gamma_{12} } )
+ O(\rho^8) \ \Bigr]dx'.   \nonumber
\end{eqnarray}
Here $V'(x)$ is  given by 
$$
 V'(x) = x'^2 + \gamma_{12} x' -{1\over 2}(\rho^2_1 +\rho^2_2),
$$ 
which is nothing but the potential (\ref{pot}) 
owing to (\ref{alpha1}) and (\ref{alpha2}). After shifting the variable $x'$ back to $x$ we  write (\ref{DV'}) 
 exressing  $\rho_i^2$ and $\gamma_i$ in terms $S_i$ and $\alpha_i$ with recourse to (\ref{rho1})$\sim$(\ref{gamma}).  
In particular  the terms of order $\rho^2$ and $\rho^4$ in (\ref{DV'}) read
\begin{eqnarray}
{\gamma_{12}\over 2}({\rho_1^2\over x'}-{\rho_2^2\over x'+\gamma_{12}})      
&=& 2\{S_1 - {2\over \alpha_{12}^3}S_1S_2- {30 \over \alpha_{12}^6}S_1S_2(S_1-S_2)\} {1\over x-\alpha_1}   \nonumber \\
&+& 2\{S_2 + {2\over \alpha_{12}^3}S_1S_2 + {30 \over \alpha_{12}^6}S_1S_2(S_1-S_2) \}{1\over x-\alpha_2}    \nonumber \\
&-& 4 \{ {1 \over \alpha_{12}^2}S_1(S_1-S_2) + 
 {2\over \alpha_{12}^5}S_1(3S_1^2 -9S_1S_2 + 4S_2^2) \} {1\over (x-\alpha_1)^2}  \nonumber   \\
&+& 4\{ {1\over \alpha_{12}^2}S_2(S_1-S_2) + 
 {2\over \alpha_{12}^5}S_2(4S_1^2 -9S_1S_2 + 3S_2^2) \} {1\over (x-\alpha_2)^2}
\nonumber \\
&+& {8 \over \alpha_{12}^4 }{S_1(S_1-S_2)^2 \over (x-\alpha_1)^3}
+ {8 \over \alpha_{12}^4 }{S_2(S_1-S_2)^2\over (x-\alpha_2)^3},    \nonumber \\
{1\over 8}\{{\rho_1^4\over x'^2}+{\rho_2^4\over (x'+\gamma_{12})^2}\} 
&=& 2 \{ {S_1^2\over \alpha_{12}^2} + 
 {4\over \alpha_{12}^5}S_1(2S_1^2 -3S_1S_2 ) \}{1\over (x-\alpha_1)^2}   \nonumber \\
&+& 2 \{ {S_2^2\over \alpha_{12}^2} + 
 {4\over \alpha_{12}^5}S_2( 3S_1S_2-2S_2^2 ) \} {1\over (x-\alpha_2)^2}  \nonumber \\
&-&{8 \over \alpha_{12}^4 }{S_1^2(S_1-S_2)\over (x-\alpha_1)^3}
+{8 \over \alpha_{12}^4 }{S_2(S_1-S_2)\over (x-\alpha_2)^3}, \nonumber \\
{\gamma_{12}\over 8}\{{\rho_1^4\over x'^3}-{\rho_2^4\over (x'+\gamma_{12})^3}\}
&=& 2\{{S_1^2\over  \alpha_{12}} + {4 \over \alpha_{12}^4 }S_1^2(S_1-2S_2)\}
{1\over (x-\alpha_1)^3} \nonumber \\
&-& 2\{{S_2^2\over  \alpha_{12}} + {4 \over \alpha_{12}^4 }S_2^2(2S_1-S_2)\} 
{1\over (x-\alpha_2)^3}  \nonumber \\
&-&{12 \over \alpha_{12}^3 }{S_1^2(S_1-S_2)\over (x-\alpha_1)^4}  
-{12 \over \alpha_{12}^3 }{S_2^2(S_1-S_2)\over (x-\alpha_2)^4}, \nonumber \\
{\rho_1^2\rho_2^2\over 4\gamma_{12}}({1\over x'}-{1\over x'+\gamma_{12}} )
&=& -4\{ {S_1S_2\over \alpha_{12}^3 } 
+ {14\over \alpha_{12}^6}S_1S_2(S_1-S_2)  \}({1\over x-\alpha_1}-{1\over x-\alpha_2})  \nonumber \\
&+&  {8 \over  \alpha_{12}^5}S_1S_2(S_1-S_2)\{{1\over (x-\alpha_1)^2}+{1\over (x-\alpha_2)^2}\}.  \nonumber
\end{eqnarray}
With these  the disc amplitude (\ref{disc}) is evaluated as
\begin{eqnarray}
F_1(S) &=& -\sum_{I=1}^{N_f}\Bigl[-S_1\log {m_I+\alpha_1\over \Lambda} -S_2\log {m_I+\alpha_2\over \Lambda}   \nonumber \\
&+&\Bigl( {1\over \alpha_{12}^2}(S_1^2-2S_1S_2) 
  +{2\over \alpha_{12}^5}(2S_1^3 -9S_1^2S_2 +6S_1S_2^2) \Bigr) {1\over m_I+\alpha_1} \nonumber \\
&+&\Bigl( {1\over \alpha_{12}^2}(S_2^2-2S_1S_2) 
  -{2\over \alpha_{12}^5}(2S_2^3 -9S_1^2S_2 +6S_1^2S_2) \Bigr) {1\over m_I+\alpha_2} \nonumber \\
&+&\Bigl( {S_1^2\over 2\alpha_{12}} 
 +{1\over \alpha_{12}^4}(2S_1^3-5S_1^2S_2+2S_1S_2^2) \Bigr){1\over ( m_I+\alpha_1)^2}  \label{F1} \\
&+& \Bigl( -{S_1^2\over 2\alpha_{12}} 
 +{1\over \alpha_{12}^4}(2S_2^3-5S_1S_2^2+2S_1^2S_2) \Bigr){1\over ( m_I+\alpha_2)^2}\nonumber \\
&+&{2\over 3\alpha_{12}^3}\Bigl\{ (2S_1^3-3S_1^2S_2) {1\over ( m_I+\alpha_1)^3}
 - (2S_2^3-3S_1S_2^2){1\over ( m_I+\alpha_2)^3}\Bigr\} \nonumber \\
&+& {1\over 2\alpha_{12}^2}\Bigl\{ {S_1^3\over (m_I+\alpha_1)^4}
 + {S_2^3\over (m_I+\alpha_2)^4} \Bigr\} +O(S^4) \Bigr].    \nonumber
\end{eqnarray}
 The $\log$-terms linear in S is a non-perturbative part which may be absorbed by the last term in the effective action (\ref{effective}). 
The terms of order $S^2$  are in complete agreement with the result (\ref{amp}) by the ${1\over N}$ expansion.

\vspace{0.5cm}

\section{The $n$-cut Solution}

We shall generalize the calculation for the $2$-cut solution to that of the $n$-cut solution.  That is, taking the matrix model in which the scalar potential is given by
\begin{eqnarray}
V'(x) = (x-\alpha_1)(x-\alpha_2)\cdots(x-\alpha_n). 
\end{eqnarray}
we consider  the DV differential 
\begin{eqnarray}
ydx = \sqrt{\mathop{\prod}_{i=1}^n \Bigl((x-\gamma_i)^2- \rho^2_i\Bigr)}dx,
  \label{DV''}
\end{eqnarray}
on the Riemann surface with $n$ branch-cuts.  Again we calculate the disc amplitude following  Step 1 to  Step 5.

\noindent
\underline {Step 1 and Step 2}

 The inversion formula $\rho_i^2(S,\gamma)$ was given in the expansion form\cite{Ito}
\begin{eqnarray}
 \rho_i^2 = {4S_i\over \Delta_i(\gamma)}\Bigl(1-{1\over 2\Delta_i(\gamma)} \sum_{j\ne k\atop j,k \ne i}{S_i\over \gamma_{ij}\gamma_{ik}} + 2\sum_{j\ne i} {S_j\over \gamma_{ij}^2\Delta_j(\gamma)}\Bigr) + O(S^3),  \label{rhoexpan}
\end{eqnarray}
in which  
\begin{eqnarray}
 \Delta_i(\gamma) = \prod_{i\ne j}^n  \gamma_{ij}, \quad\quad\quad 
\gamma_{ij}=\gamma_i-\gamma_j. \nonumber
\end{eqnarray}
As  it will be clear, the expansion up to  order $S^2$ is sufficient for the calculation of the disc amplitude to order $S^3$. 

\noindent
\underline{Step 3}

We expand $y$, given by (\ref{SPE}) and  (\ref{DV''}), at $x=\infty$. 
Comparing the polynomial part of both expansions  yields\cite{Ito} 
\begin{eqnarray}
 2\sum_{i=1}^n \alpha_i^l= \sum_{i=1}^n \Bigl[(\gamma_i+\rho_i)^l + 
 (\gamma_i-\rho_i)^l \Bigr], \quad\quad 1\le l \le n.    \label{sympoli}
\end{eqnarray}
We shall express $\gamma_i$ in terms of $\rho$ and $\alpha$ by solving this set of equations recursively. 
First of all expand (\ref{sympoli}) in series of $\rho^2$ as
\begin{eqnarray}
 \sum_{i=1}^n \alpha_i^l = \sum_{i=1}^n \Bigl[\gamma_i^l + {l\choose 2}\gamma_i^{l-2}\rho_i^2 + {l\choose 4} \gamma_i^{l-4}\rho_i^4 + O(\rho^6) \Bigr].
 \label{sympoliex}
\end{eqnarray}
Assume that $\gamma_i$ is further expanded in the form
\begin{eqnarray}
\gamma_i = \alpha_i + \sigma_i^{(2)} + \sigma_i^{(4)} + O(\rho^6),
  \label{gammainverse}
\end{eqnarray}
with $\sigma_i^{(n)}$ the term of order $\rho^n$. Putting  this in (\ref{sympoliex}) we get the recursion relations for $\sigma_i^{(n)}$
\begin{eqnarray}
\sum_{i=1}^n \Bigl[\alpha_i^{l-1}\sigma_i^{(2)} &+& {1\over 2}(l-1) \alpha_i^{l-2}\rho_i^2\Bigr] =0,  \label{recur1} \\
 \sum_{i=1}^n \Bigl[ \alpha_i^{l-1}\sigma_i^{(4)} &+& {1\over 2} (l-1)\alpha_i^{l-2}(\sigma_i^{(2)})^2 + {1\over 2}(l-1)(l-2)\alpha_i^{l-3}\sigma_i^{(2)}\rho_i^2 
 \nonumber \\
&+&  {1\over 4!}(l-1)(l-2)(l-3) \alpha_i^{l-4}\rho_i^4 \Bigr] = 0  \label{recur2}, 
\end{eqnarray}
and so on. Note that the identity
\begin{eqnarray}
 {P^{[i]}(\alpha_j)\over \Delta_i(\alpha)} = \delta_{ij},
\label{identity}
\end{eqnarray}
for the quantity 
\begin{eqnarray}
  P^{[i]}(x) = {V'(x) \over x-\alpha_i} = \prod_{l\ne i}^n (x-\alpha_l).
 \label{P}
\end{eqnarray}
Using the expansion
$$
P^{[i]}(x) = \sum_{l=1}^n c_l^{[i]}(\alpha) x^{l-1},
$$
(\ref{identity}) becomes 
$$
{1\over \Delta_i(\alpha)}\sum_{l=1}^n 
c_l^{[i]}(\alpha) \alpha_j^{l-1}  = \delta_{ij}.
$$
By means of this identity the recursion relations (\ref{recur1}) and (\ref{recur2}) 
can be solved for $\sigma_i^{(n)}$ as\begin{eqnarray}
 \sigma_i^{(2)} &=& -{1\over 2\Delta_i(\alpha)} \sum_{j=1}^n P^{[i]}{'}(\alpha_j)\rho_j^2,  \label{sigma1}  \\
\sigma_i^{(4)} &=&  -{1\over 2\Delta_i(\alpha)} \sum_{j=1}^n P^{[i]}{'}(\alpha_j) (\sigma_i^{(2)})^2 -{1\over 2\Delta_i(\alpha)} \sum_{j=1}^n P^{[i]}{''}(\alpha_j)\sigma_i^{(2)}\rho_j^2  \nonumber \\
&\quad& \hspace{1cm} -{1\over 24\Delta_i(\alpha)} \sum_{j=1}^n P^{[i]}{'''}(\alpha_j)\rho_j^4. \label{sigma2} 
\end{eqnarray}
Here $P^{[i]}{'}(x), P^{[i]}{''}(x), P^{[i]}{'''}(x) $ are  derivatives of the quantity  (\ref{P}) with respect to $x$. Putting (\ref{sigma1}) and (\ref{sigma2}) in (\ref{gammainverse}) we  obtain  the formula generalizing (\ref{gamma1}) and (\ref{gamma}) for the $n$-cut solution. 

\noindent
\underline {Step 4}

Expanding  $y$ given by the DV differential (\ref{DV''}) in series of $\rho^2$  we get 
\begin{eqnarray}
y&=& -\sum_{i=1}^n \Bigl[{1\over 2}{\rho_i^2\over x-\gamma_i}
+{1\over 8}{\rho_i^4\over (x-\gamma_i)^3} +{1\over 16}{\rho_i^6\over (x-\gamma_i)^5} \Bigr]\tilde P^{[i]}(x)  \nonumber \\
&\quad& +\sum_{i\ne j}^n \Bigl[{1\over 8}{\rho_i^2\rho_j^2\over (x-\gamma_i)(x-\gamma_j)}
 + {1\over 16}{\rho_i^2\rho_j^4  \over (x-\gamma_i)(x-\gamma_j)^3}\Bigr]\tilde P^{[ij]}(x)  \label{expy} \\
&\quad&-{1\over 48}\sum_{i\ne j\ne k}^n {\rho_i^2\rho_j^2\rho_k^2\over (x-\gamma_i)(x-\gamma_j)(x-\gamma_k)}\tilde P^{[ijk]}(x) + O(\rho^8), \nonumber
\end{eqnarray}
with
\begin{eqnarray}
  \tilde P^{[i]}(x) &=& \prod_{l\ne i}^n (x-\gamma_l), \quad\quad\quad
 \tilde P^{[ij]}(x) = \prod_{l\ne i,j}^n (x-\gamma_l),  \nonumber \\
\tilde P^{[ijk]}(x) &=& \prod_{l\ne i,j,k}^n (x-\gamma_l).  \nonumber
\end{eqnarray}
When $\tilde P^{[i]}(x), \tilde P^{[ij]}(x)$ and $\tilde P^{[ijk]}(x)$ are expanded at $x= \gamma_i, \gamma_j, \gamma_k$  in Taylor series, it takes the form 
\begin{eqnarray}
  y =  V'(x) +  \sum_{\nu=1}^5 \sum_{i=1}^n{a_{\nu i} (\rho^2,\gamma)\over (x-\gamma_i)^\nu} + O(\rho^8),   \nonumber
\end{eqnarray}
with 
\begin{eqnarray}
a_{1i}(\rho^2,\gamma) &=&  -{\rho_i^6\over 384}\tilde P^{[i]}{''''}(\gamma_i)  - {\rho_i^4\over 16}\tilde P^{[i]}{''}(\gamma_i)  -{\rho_i^2\over 2}\tilde P^{[i]}(\gamma_i)
  \nonumber \\ 
&+& \sum_{j\ne i}^n \Bigl[- {\rho_i^4\rho_j^2\over 16\gamma_{ij}^2}\tilde P^{[ij]}{'}(\gamma_i) + {\rho_i^4\rho_j^2\over 32\gamma_{ij}}\tilde P^{[ij]}{''}(\gamma_i) \nonumber \\ 
&\quad& \quad +{\rho_i^2\rho_j^2(\rho_i^2+ \rho_j^2)\over 16\gamma_{ij}^3}\tilde P^{[ij]}(\gamma_i) +{\rho_i^2\rho_j^2\over 4\gamma_{ij}}\tilde P^{[ij]}(\gamma_i) \Bigr]
\nonumber \\
 &-& \sum_{j\ne k \atop j,k\ne i}^n {\rho_i^2\rho_j^2\rho_k^2\over 24}({1\over \gamma_{ij}\gamma_{ik}} + {1\over \gamma_{ij}\gamma_{jk}})\tilde P^{[ijk]}(\gamma_i),   \label{yexpan} \\ 
a_{2i}(\rho^2,\gamma) &=&  -{\rho_i^6\over 96}\tilde P^{[i]}{'''}(\gamma_i)
-{\rho_i^4\over 8}\tilde P^{[i]}{'}(\gamma_i)    
- \sum_{j\ne i}^n{\rho_i^4\rho_j^2\over 16\gamma_{ij}^2}\tilde P^{[ij]}(\gamma_i)
+ \sum_{j\ne i}^n {\rho_i^4\rho_j^2\over 16\gamma_{ij}}\tilde P^{[ij]}{'}(\gamma_i),
 \nonumber \\ 
a_{3i}(\rho^2,\gamma) &=&  -{\rho_i^6\over 32}\tilde P^{[i]}{''}(\gamma_i)
 - {\rho_i^4\over 8}\tilde P^{[i]}(\gamma_i) 
 + \sum_{j\ne i}^n{\rho_i^4\rho_j^2\over 16\gamma_{ij}}\tilde P^{[ij]}(\gamma_i),
     \nonumber \\ 
a_{4i}(\rho^2,\gamma) &=& -{\rho_i^6\over 16}\tilde P^{[i]}{'}(\gamma_i),       \nonumber \\ 
a_{5i}(\rho^2,\gamma) &=&  -{\rho_i^6\over 16}\tilde P^{[i]}(\gamma_i).  
     \nonumber
\end{eqnarray}
The polynomial part $V'(x)$ was determined  by the fact that $y$ given by  (\ref{expy}) may be written in the form (\ref{SPE}) as well.

\noindent
\underline {Step 5}

Finally we use the inversion formulae for $\rho_i^2$ and $\gamma_i$, given by (\ref{rhoexpan}) and (\ref{gammainverse}) respectively, to find $y$ in the form to order $S^3$
\begin{eqnarray}
y = V'(x) +  \sum_{\nu=1}^5\sum_{i=1}^n {b_{\nu i} (S)\over (x-\alpha_i)^\nu} +O(S^4),
 \label{yexp'}
\end{eqnarray}
with
\begin{eqnarray}
b_{1i}(S)    &=&  -2S_i,           \nonumber \\ 
b_{2i}(S)  &=& \sum_{j\ne i}^n\Bigl[{2S_i^2-4S_iS_j\over \Delta_i(\alpha)\alpha_{ij}}  + 
4{S_i^3 - 6S_i^2S_j+3S_iS_j^2 \over \Delta_i(\alpha)^2\alpha_{ij}^3}
-4{S_i^3 - 3S_i^2S_j+3S_iS_j^2 \over \Delta_i(\alpha)\Delta_j(\alpha)\alpha_{ij}^3}\Bigr]   \nonumber \\ 
&+& \sum_{j\ne k \atop j,k\ne i}^n \Bigl[
4{S_i^3-8S_i^2S_j+S_iS_j^2+2S_i^2S_k  +4S_iS_jS_k\over \Delta_i(\alpha)^2\alpha_{ij}^2 \alpha_{ik}}   
 -4{S_i^3 -4S_i^2S_j+2S_i^2S_k  \over \Delta_i(\alpha)\Delta_j(\alpha)\alpha_{ij}^2 \alpha_{ik}} \nonumber \\ 
&\quad& \quad +4{3S_iS_j^2-4S_i^2S_j+4S_i^2S_k -4S_iS_jS_k \over \Delta_i(\alpha)\Delta_j(\alpha)\alpha_{ij}^2 \alpha_{jk}}   
-4{S_iS_j^2+S_iS_k^2-2S_iS_jS_k \over \Delta_i(\alpha)\Delta_j(\alpha)\alpha_{ij} \alpha_{jk}^2}   \nonumber \\ 
&\quad& \quad +8{S_iS_j^2-2S_iS_jS_k \over \Delta_i(\alpha)\Delta_k(\alpha)\alpha_{ij} \alpha_{jk}^2}  
-4{S_i^2S_j +S_i^2S_k-2S_iS_jS_k \over \Delta_i(\alpha)\Delta_j(\alpha)\alpha_{ij}\alpha_{ik} \alpha_{jk}} \Bigr]  \nonumber \\ 
&+& \sum_{j\ne k \ne l \atop j,k,l\ne i}^n \Bigl[ 
{{2\over 3}S_i^3 +10S_i^2S_j - 4S_i^2S_k -10S_i^2S_l + 4S_iS_kS_l\over \Delta_i(\alpha)^2\alpha_{ij}\alpha_{ik}\alpha_{il}}  \nonumber \\
&\quad& \quad +{2S_i^2S_j -{16\over 3}S_iS_j^2 - 4S_iS_jS_k +12S_iS_jS_l 
-4S_iS_kS_l \over \Delta_i(\alpha)\Delta_j(\alpha)\alpha_{ij}\alpha_{jk}\alpha_{jl}} \nonumber \\ 
&\quad& \quad -4{S_i^2S_k-S_i^2S_l \over \Delta_i(\alpha)\Delta_k(\alpha)\alpha_{ij}\alpha_{ik}\alpha_{kl}}
+8{S_iS_jS_k-S_iS_jS_l  \over \Delta_i(\alpha)\Delta_k(\alpha)\alpha_{ij}\alpha_{jk}\alpha_{kl}} \Bigr]   
+ O(S^4),    \nonumber  \\
b_{3i}(S) &=& -{2S_i^2\over \Delta_i(\alpha)}    
 - \sum_{j\ne i}^n \Bigl[{4S_i(S_i-S_j)(S_i-2S_j)\over \Delta_i(\alpha)^2\alpha_{ij}^2} -{4S_i^2(S_i-2S_j)\over \Delta_i(\alpha)\Delta_j(\alpha)\alpha_{ij}^2}\Bigr] \nonumber \\ 
&\quad& -\sum_{j\ne k\atop j,k\ne i}^n \Bigl[ {4S_i(S_i-S_j)(S_i-2S_k)\over \Delta_i(\alpha)^2\alpha_{ij}\alpha_{ik}} 
-{4S_i^2(S_j-S_k)\over \Delta_i(\alpha)\Delta_j(\alpha)\alpha_{ij}\alpha_{jk}}
\Bigr] + O(S^4),   \nonumber \\ 
b_{4i}(S)  &=&  \sum_{j\ne i}^n{4S_i^2(2S_i-3S_j)\over \Delta_i(\alpha)^2\alpha_{ij}}+ O(S^4),    \nonumber \\ 
b_{5i} (S)  &=&  -{4S_i^3\over \Delta_i(\alpha)^2} +O(S^4).   \nonumber  
\end{eqnarray}
Here the coefficient $b_{1i}(S)$ is the exact result obtained  
 by comparing the term tending to ${1\over x}$ at $x=\infty$ in (\ref{SPE}) and (\ref{expy}). We would like to remark that the inversion formula (\ref{rhoexpan}) for $\rho_i^2$ to order $S^2$ is sufficient to evaluate the other coefficients  $b_{\nu i}(S)$'s to order $S^3$. Having explicitly known $y$ in the expansion form  (\ref{yexp'}) it is easy to evaluate the disc amplitude (\ref{disc}) for the $n$-cut solution. We then see that for $n=2$ it exactly reproduces the disc amplitude given by (\ref{F1}). Moreover we note that the evaluation of  (\ref{disc})  to  order $S^2$ gives the  disc amplitude (\ref{amp})   in the extended form  to the generic $n$, which was obtained by the $1\over N$ expansion in \cite{Sch}.

\vspace{0.5cm}

\section{Conclusions}

To the author's knowledge the disc amplitude for the multi-cut solution was calculated only the diagramatic expansion in the literature. The result  was merely reported to order $S^2$ for the generic $n$-cut solution\cite{Sch}. The calculation to higher orders can be hardly done, because the diagramatic expansion becomes  complicate as the degree of the polynomial scalar potential increases. To confront with this situation we exploited the alternative method, {\it i.e.}, the complex analysis of the (hyper)\-elliptic curve. In section 4 
the method  was worked out to order $S^3$ for the disc amplitude for the $2$-cut solution. Dealing with the elliptic curve it is relatively easy to extend the calculation to higher orders.  In section 5 the argument was generalized for the generic $n$-cut solution.  By using  a more systematic  algorithm  than  in section 4, the disc amplitude for the $n$-cut solution was calculated  up to order $S^3$. There we have given  the inversion formula for $\gamma_i$   to order $\rho^6$. We would like remark that it was sufficient to give the formula   to order $\rho^4$  for the calculation of the planar amplitude to order $S^3$\cite{Ito}. It was checked that the disc amplitude thus obtained  indeed reproduces the one for the $2$-cut solution obtained by the ${1\over N}$ expansion in \cite{Sch}. 
The algorithm in this paper  allows us to generalize the calculation of the disc amplitude for the $n$-cut solution to higher orders. 

\vspace{1cm}

\noindent
Note Added:

After posting the paper on the arXiv, the author was told that the disc amplitude for the $n$-cut solution was calculated to order $S^3$ by a different method in \cite{Reino}. 
But the result (4.16) in \cite{Reino} is erroneous. After the correction it reads 
\begin{eqnarray}
{\cal F}^{\chi=1}_3 &=& \sum_{f=1}^{N_f}\sum_{i=1}^n\Bigl[ -{1\over 2}{S_i^3\over R_i^2 e_{if}^4} - \sum_{j\ne i}^n\Bigl\{
   {2\over 3}{S_i^2(2S_i-3S_j)\over R_i^2 e_{ij}e_{if}^3}+ {2S_i^3 -5S_i^2S_j + 2S_iS_j^2 \over R_i^2 e_{ij}^2 e_{if}^2}
   \nonumber \\
&\quad&  +{6S_i^3 -26S_i^2S_j+14S_iS_j^2\over R_i^2e_{ij}^3 e_{if}} +
  {2S_i^3 -8S_i^2S_j+2S_iS_j^2\over R_i R_je_{ij}^3 e_{if}}   \Bigr\} 
+\sum_{j\ne i}^n\sum_{k\ne j,i}^n\Bigl\{\cdots \Bigr\}  \Bigr],   \nonumber 
\end{eqnarray}
by extracting the single sum contribution $\sum_{j\ne i}\{\cdots \}$  from 
$\sum_{j\ne i}\sum_{k\ne i}$ and $\sum_{j\ne i}$  $\sum_{k\ne j}$. 
Then this  agrees with  the disc amplitude given by (\ref{disc}) with (\ref{yexp'}) in our paper, up to the double sum contribution 
$\sum_{i=1}\sum_{j\ne i}\{\cdots\}$. 
To check this,  note that with the notation in \cite{Reino}
\begin{eqnarray}
 \sum_{j\ne i}^n{S_j^{\ N-1}\over e_{ij}^{\ M }R_j} = -\sum_{j\ne i} {S_j^{\ N-1}\over e_{ij}^{\ M }R_i} + \sum_{j\ne i}^n\sum_{k\ne j,i}^n\Bigl\{\cdots\Bigr\}, 
\quad\quad\quad M,N = 1,2,\cdots, \nonumber
\end{eqnarray}
which follows from 
\begin{eqnarray}
 \sum_{i=1}^n\oint_{e_i} {dx\over (x-e_i)^M W'(x)} = 0, \quad\quad
\sum_{l=1}^n \sum_{i=1}^n \oint_{e_i} {S_l^{\ N-1}dx\over (x-e_l) (x-e_i)^M W'(x)} = 0.      \nonumber
\end{eqnarray}
It is more involved to check the consistency between both results for the  triple sum contribution $\sum_{i=1}\sum_{j\ne i}\sum_{k\ne j,i}\{\cdots \}$. 
The above calculation uses the resolvent and the algorism developed in \cite{Morosov, Reino}. It is simple, but allows us to evaluate  merely counter integrals of the DV differential. On the other hand the calculation in our paper uses the other algorism \cite{Ito}. It allows us to argue on the DV differential itself, expressing the branch points of the Riemann surface  as functions of $S_i$'s. The latter algorism is important to study the Whitham hierarchy of the DV theory.

\vspace{2cm}
\noindent

\acknowledgments

\noindent
The work was supported in part  by the Grant-in-Aid for Scientific Research No.
13135212.

\end{document}